\documentclass[twoside,11pt]{article}

\usepackage{jmlr2e}
\usepackage{multirow}
\usepackage{graphicx}
\usepackage{boldline}
\usepackage{subcaption}
\usepackage{amsmath}
\usepackage{pdfpages}

\graphicspath{ {./images/} }

\DeclareMathOperator*{\ent}{\textit{n}}

\jmlrheading{}{}{}{}{}{}{}

\ShortHeadings{KATRec}{Amjadi, Mohseni-Taheri and Tulabandhula}
\firstpageno{1}

\begin{document}

\title{KATRec: Knowledge Aware aTtentive Sequential Recommendations}

\author{\name Mehrnaz Amjadi$^*$ \email mamjad2@uic.edu 
       \AND
       \name Seyed Danial Mohseni Taheri$^*$ \email smohse3@uic.edu 
       \AND
       Theja Tulabandhula \email tt@theja.org{ } \\
       \addr Information and Decision Sciences\\
       University of Illinois\\
       Chicago, IL 60607, USA}

\editor{}

\maketitle

\begin{abstract}%
  Sequential recommendation systems model dynamic preferences of users based on their historical interactions with platforms. Despite recent progress, modeling short-term and long-term behavior of users in such systems is nontrivial and challenging. To address this, we present a solution enhanced by a knowledge graph called KATRec (Knowledge Aware aTtentive sequential Recommendations). KATRec learns the short and long-term interests of users by modeling their sequence of interacted items and leveraging pre-existing side information through a  knowledge graph attention network. Our novel knowledge graph-enhanced sequential recommender contains item multi-relations at the entity-level and users' dynamic sequences at the item-level. KATRec improves item representation learning by considering higher-order connections and incorporating them in user preference representation while recommending the next item. Experiments on three public datasets show that KATRec outperforms state-of-the-art recommendation models and demonstrates the importance of modeling both temporal and side information to achieve high-quality recommendations.
\end{abstract}


\section{Introduction}
With the exploding growth of online platforms in recent years, recommendation systems have become an essential component in elevating user  engagement levels and thus have taken a central role in business success. Many services leverage historical data of users and their interactions with their service (e.g., an app or a website) to  personalize recommendations. Such recommendation systems are increasingly popular in various domains including: e-commerce, social media, search engines, content portals, and online publishing platforms.

In this work, we focus on exploiting the sequential behavior of users in order to predict their upcoming interactions. Existing sequential recommender designs, e.g., Markov chains, recurrent neural networks (RNN), graph convolutional neural networks (GCN), and self-attention based models (to name a few), primarily focus on various ways to model such historical data. However, these sequential models tend to disregard the relationship between items. In particular, platforms usually have access to two types of information that can be valuable for recommendations: (i) interactions of users and the service, which may evolve over time, and (ii)  side information about users, items, and other auxiliary components. Recommender systems can generate more relevant content by taking advantage of item relations that are hard to elicit from interaction sequences of users. Such side information about the items can be based on higher-order item-entity connections and co-occurrence patterns, which can provide implicit information about related items (for instance, a mouse and a laptop).

Efficient exploitation of both temporal data of users and side information is the primary gap this paper attempts to fill. While there are many well-performing solutions in the literature, they do not effectively capture both types of information. For instance, models such as BERT4Rec, SASRec, and GRU4Rec \citep{sun2019bert4rec, kang2018self, hidasi2015session} heavily focus on the temporal aspect by encoding user behavior sequences. Another stream of literature focuses on graph structure to capture item relations and side information, for example: TransE \citep{bordes2013translating}, TransH \citep{wang2014knowledge}, and TransR \cite{lin2015learning}. In this work, we build on both these prior works and provide a novel way to integrate them. We follow this up by systematically exploring the importance of capturing both types of information on recommendation quality.

\begin{figure}[h]
\centering
\includegraphics[  page=1,
  width=0.7 \textwidth,
  height=\textheight,
  keepaspectratio,
  trim=0 0 0 20pt,]{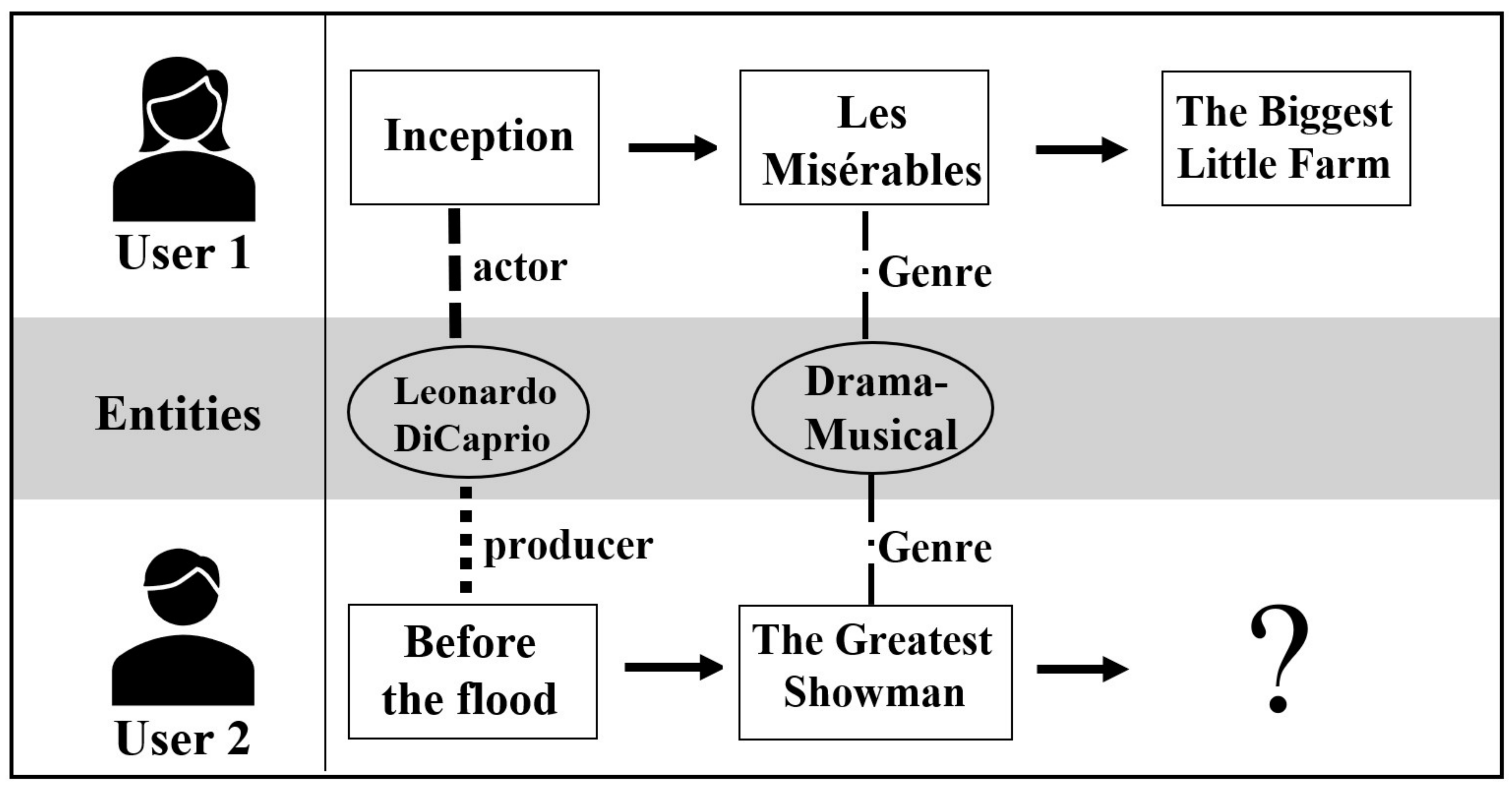}\caption{Sequential interactions of two users with the systems. While users interact with different items, their collaborative signal can be detected via shared entities.}\label{illustration}
\end{figure}

To capture the short-term preferences, long-term interests, and item-item relations, we propose the Knowledge Aware aTtentive Sequential Recommendations (KATRec) system. Our recommendation system consists of two modules: (i) a bidirectional transformer, which captures sequential interests by considering the inter-dependencies among items at any temporal distance, and (ii) a knowledge graph attention network that models higher-order user-item and item-item relations. The user-item relations are based on the interactions of users with items, e.g., click, purchase, view, etc., and capture \emph{collaborative information}. The item-item relations capture the \emph{semantic relatedness} among items based on their shared entities (e.g., movies with the same actor or genre). The importance of capturing such information while making sequential recommendations has been previously discussed in \citet{ji2019sequential, ma2019memory, zhang2020adaptive} to name a few. In addition to first-order relations, sequential recommendations generated by KATRec can use higher-order connections and relations among items (e.g., an individual can be an actor in a movie and a producer in another). Figure~\ref{illustration} illustrates two sequences of user interactions. A traditional sequential recommendation system models these two users differently, as they interact with different movies, although these users show similar interests to movies with shared entities (genre, actor, producer, etc.). Therefore, incorporating information using a knowledge graph can enhance their representations, and consequently improve recommendation performance \citet{KGAT19, zhao2019kb4rec}.

This work proposes a novel deep neural network architecture incorporating both sequential behavior of users and side information about items. Our proposed structure captures the temporal information using a sequential attention mechanism and spatial information via a knowledge graph attention mechanism. To summarize, the key contributions of this work are:
\begin{itemize}
    \item KATRec builds on a knowledge graph neural network that captures multi-relationships between items by tying them together using the underlying entities. This allows for better representations of items and enhances the recommendation performance.
    \item KATRec models the short-term and long-term user preferences by adaptively aggregating dynamic interactions and item-item multi-relations through a gating mechanism. This mechanism can significantly alleviate the sparsity in both user sequences and item relationships.
    \item KATRec captures co-occurrence information and collaborative signals by leveraging attention mechanisms in the knowledge graph, which ultimately impact the attention weights in the bidirectional transformer module and the overall recommendations.
    \item We conduct experiments to evaluate the impact of different components on the performance of KATRec, and show that it outperforms state-of-the-art baselines on three public datasets.
\end{itemize}

The paper is organized as follows: in Section \ref{lit}, we review the relevant prior literature. Section \ref{model} defines the problem and presents the new architecture KATRec. We conduct a detailed comparison of the performance of KATRec with multiple competitive baselines in Section~\ref{experiment}. Section~\ref{conclusion} concludes with some pointers to future directions.

\section{Literature Review}\label{lit}

\subsection{General Recommendations}
Early work on recommendation systems uses collaborative filtering (CF) to make recommendations based on common interests between users \citep{koren2008factorization}, which can be modeled using either explicit or implicit ratings \citep{hu2008collaborative}. CF based models usually suffer from the sparsity of the user-item interactions, which makes them susceptible to the cold start problem. To address this issue, newer variants of matrix factorization, including point-wise and pair-wise methods, have been proposed \citep{rendle2012bpr, he2016fusing}. 

Recent advances in deep neural networks (DNNs) have resulted in the development of recommendation models that have higher flexibility in learning user and item representations \citep{zhang2019deep}. Among several recent works in this area, \cite{wu2016collaborative} and \cite{he2017neural} use conventional matrix factorization with DNNs to predict ratings using autoencoders, and also predict user preferences using certain multi-layer perceptron architectures.  

\subsection{Sequential Recommendations}
Given a user's temporally structured historical interaction logs, sequential recommendations predict the next item that a user is likely to adopt. Early works incorporated Markov chains to capture the underlying temporal behavior. For example, \citep{rendle2010factorizing, he2016fusing} look at the most recently interacted item(s) to predict the next item to recommend. Newer methods have incorporated higher-order Markov chains to capture a better representation of user behavior. For instance, the Convolutional Sequence Model (Caser) in \cite{tang2018personalized} learns the sequential patterns of a user via an embedding matrix of a fixed size that captures the previous $L$ items, using it as an image input and leveraging convolutional filters. A collection of works has similarly adopted RNNs to encode temporal user preferences into a context vector~\citep{hidasi2015session, donkers2017sequential}. 
The attention mechanism has recently been used as a key modeling component for time series data in natural language processing, including in architectures such as the transformer \citep{vaswani2017attention} and BERT \citep{devlin2018bert}. Inspired by these initial works, \cite{kang2018self} proposes a two-layer transformer decoder, namely SASRec, to capture sequential user behavior. \cite{sun2019bert4rec} encodes user preferences with a bidirectional model using the Cloze task setup. While all these models encode sequential preferences of users, they have limited power in capturing the relationships between items themselves and between items and users in a parsimonious way. Our proposed approach aggregates higher-order and multi-relation item connections and dynamic preferences of users to predict the next item while retaining modular interpretability.

\subsection{Knowledge-aware Recommendations}
There have been multiple prior works that focus on enhancing recommendation system performance by explicitly considering multi-relations between items \citep{qin2020survey}. These models usually leverage the structure of an associated knowledge graph and a static user-item interaction graph to exploit such higher-order relationships, typically through path selection algorithms \citep{wang2019explainable, lei2020interactive} and meta-paths heuristics \citep{huang2020meta,hu2018leveraging}. However, these approaches can be sub-optimal for the recommendation objective, and may requires a reasonably high degree of domain knowledge, making them less suitable for generating highly relevant recommendations. A recent work focuses on regularizing the recommendation objective by adding a loss that captures the knowledge graph structure \citep{wang2019knowledge,KGAT19}. Due to certain modeling choices, it is not fully clear if higher-order connectivity is being captured effectively, while retaining interpretability. In contrast to these methods, some of which are focused on generic graphs, we incorporate a knowledge graph while taking the higher-order connectivity between entities and their neighbors into account in a parsimonious way, and fuse these signals into an attention-based sequential model that captures dynamic preferences of a user.

\section{Method}\label{model}

Our goal is to provide a personalized next item recommendation for users based on their history and higher-order relations between items. In this section, we first state a formal definition of the problem, and then we elaborate on different parts of our proposed solution.

\subsection{Problem Definition and Solution Overview}

Given a set of users $\mathcal{U}$ and items $\mathcal{I}$, we have a sequence of items $\mathcal{S}^u = \{S_1^u,  \cdots, S^u_{T}\}$ that user $u$ has interacted with over $T$ time steps ($T = |\mathcal{S}^u|$). Also, we have access to side information related to items (e.g., actors, directors, and genre as shown in Figure \ref{illustration}). Based on the historical interaction sequence $\mathcal{S}^u$, our goal is to predict the item that user $u$ is most likely interested in at the next time step $T+1$.

\begin{figure}[h]
\centering
\includegraphics[width=\textwidth]{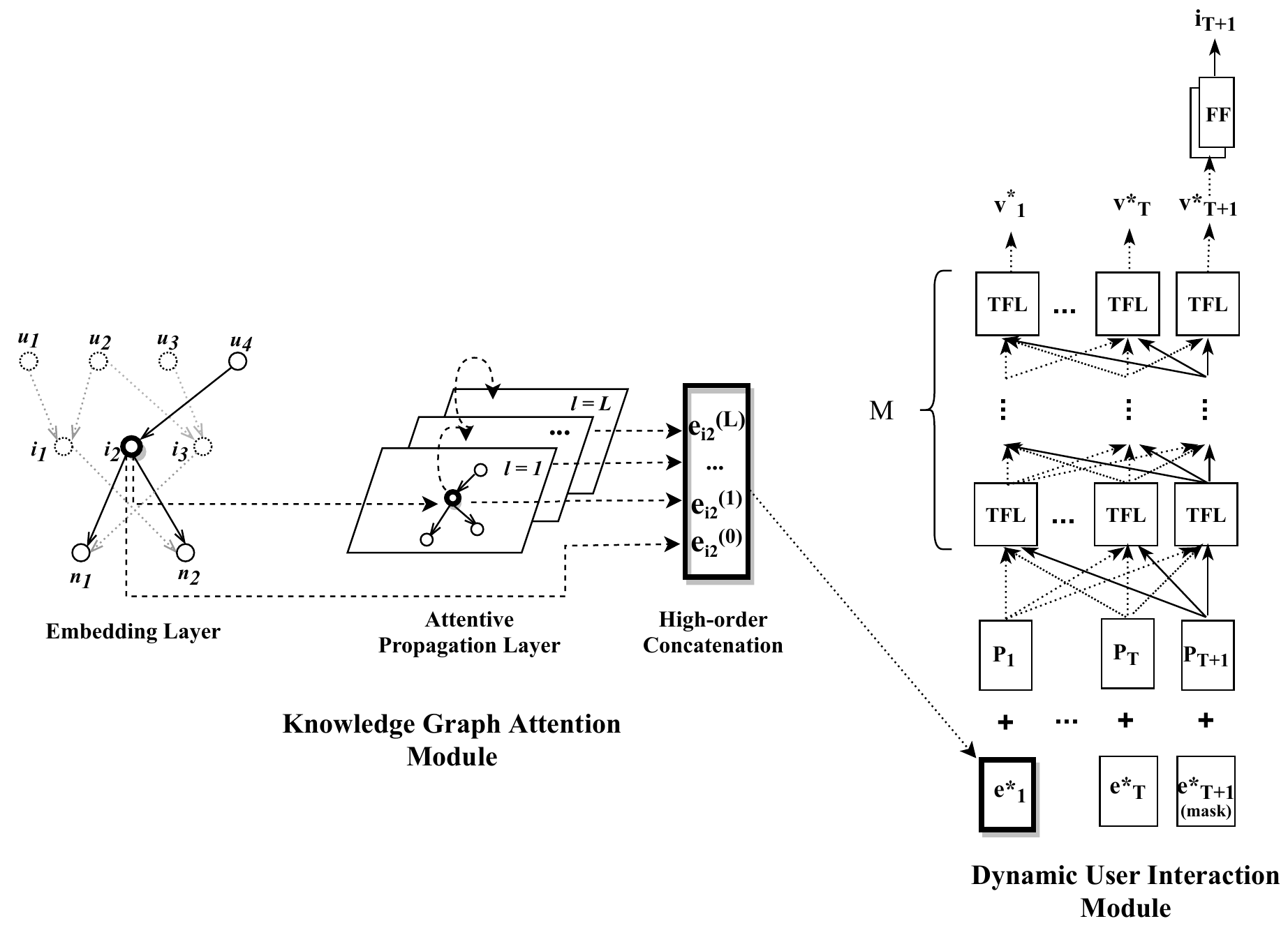}\caption{Illustration of different components of the knowledge graph attention module (left) and the dynamic user interaction module (right). First, we learn the initial item embeddings using the knowledge graph attention module. Then, we feed these item embeddings and the user's interaction sequence $\mathcal{S}^u$ through $M$ layers of the bidirectional transformer (TFL) to obtain the final item embeddings that are then used for next item recommendation.}\label{katrec}
\end{figure}

In KATRec, we build a knowledge-aware attentive sequential recommendation system to facilitate modeling of dynamic behavior of users while capturing the multi-relations between items. Specifically, our model contains two modules as shown in Figure \ref{katrec}, namely: (a) a graph neural network $\mathcal{G}$ which captures item level multi-relations, and (b) a bidirectional transformer module that incorporates item embeddings from the knowledge graph network into the representations of dynamic preferences exhibited by the users. In the following, we discuss the details of each.

\subsection{Knowledge Graph Module with Attention}\label{subsec:gat}

\begin{figure}[h]
\centering
\includegraphics[width=0.5\textwidth]{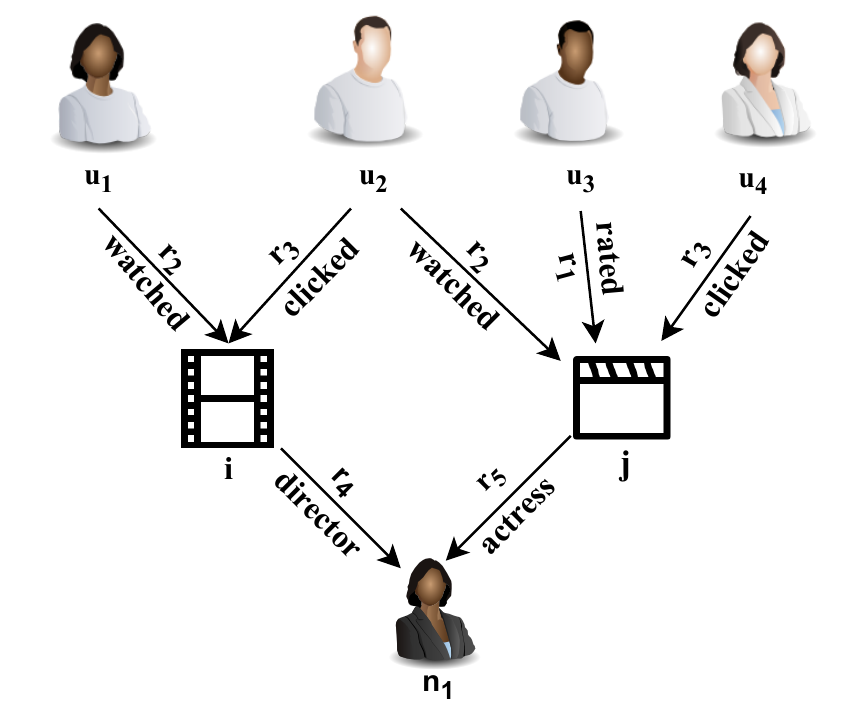}\caption{Collaborative knowledge graph relates items $i$ and $j$ through user $u_2$ and entity $\ent_1$ with different types of relations.}\label{kg_fig}
\end{figure}

This module encodes items' metadata as a unified graph to exploit the higher-order connectivity between items. The graph $\mathcal{G} = (\mathcal{E}, \mathcal{R})$, where $\mathcal{I} \subset \mathcal{E}$, incorporates entities as nodes and relationships as the edges. For instance, entities in a movie data set include the items and their side information e.g., genres, producer, actor, etc.  We use entities as building blocks that capture connections between items, and we focus on paths that start and end with items. Formally, the $K$-order connectivity between items is a path that captures a higher-order relationship between items $(i, j)$ as: $i \xrightarrow{r_1} \ent_1 \xrightarrow{r_2} \ent_2 \xrightarrow{r_3} \cdots \xrightarrow{r_K} j$, where $(i, j) \in \mathcal{I}$, entity $\ent_k \in \mathcal{E}$, and relation $r_k \in \mathcal{R}$ for $k \in \{1,2, \cdots, K\}$. In addition to item and entities in the paths, we also model the joint occurrence of commonly related items using a collaborative knowledge graph by adding users as nodes in the graph. In particular, we integrate the item-user relations $\mathcal{R}_{\mathcal{U}}$ into the knowledge graph, so in the resulting graph, the interactions of users with items are also captured. We can represent the graph by a pair of nodes and their relation as $\mathcal{G} = \{(h, r, t)|h,t \in \mathcal{E}', r \in \mathcal{R}'\}$, where $\mathcal{E}' \subset \mathcal{E} \cup \mathcal{U}$ and $\mathcal{R}' \subset \mathcal{R} \cup \mathcal{R}_{\mathcal{U}}$. Figure \ref{kg_fig} illustrates a collaborative knowledge graph where movie $i$ and movie $j$ are second order neighbors related by entity $\ent_1$ in the path $i \xrightarrow{r_4} \ent_1 \xrightarrow{r_5} j$ or related by user $u_2$ in the path $i \xrightarrow{r_3} u_2 \xrightarrow{r_2} j$.

In order to encode these relations as item embeddings, we use the TransR method \citep{lin2015learning}. TransR learns the embedding of each node and relation via the translation principle: $W^r e_h + e_r \approx W^r e_t$ ($e_h, e_t \in \mathbb{R}^d$, $e_r\in \mathbb{R}^k$, and $W^r \in \mathbb{R}^{k\times d}$), if triplet $(h, r, t)$ exists in the knowledge graph. For each triplet $(h, r, t)$ the dissimilarity score is computed using:
\begin{displaymath} s(h, r, t) = \lVert W^r e_h+e_r - W^re_t \rVert_2^2,
\end{displaymath}
The lower the score of $s(h, r, t)$ the more likely is the triplet in the KG. Following \cite{KGAT19}, we capture each relation's importance by generating attentive weights between a node and its higher-order neighbors. So, for each node $h$, we initially consider all nodes that have the first-order relation with it, i.e.,
\begin{displaymath}\mathcal{N}_{h} = \{(h', r, t)|(h', r, t)\in \mathcal{G} \textrm{ with } h'=h\}.
\end{displaymath}

The first order connectivity embedding of head node $h$ is defined as the linear combination of the embeddings of its ego network:
\begin{displaymath}
e_{\mathcal{N}_h} = \sum_{(h, r, t)\in \mathcal{N}_h} \pi(h, r, t)e_t,
\end{displaymath}
where attention factor $\pi(h, r, t)$ controls how much information from different tails can be propagated to head $h$ based on specific relations, and can be computed as follows:
\begin{displaymath}
\pi(h, r, t) = (W^r e_t)^\top \tanh(W^r e_h+e_r).
\end{displaymath}

This attention mechanism will propagate more information from closer entities in the relationship space. Then, we normalize the coefficients across all $h$'s first-order relations using a softmax function:
\begin{displaymath}\pi'(h, r, t) = \frac{\exp(\pi(h, r, t))}{\sum_{(h, r', t')\in \mathcal{N}_h}\exp(\pi(h, r', t'))}.
\end{displaymath}

We update the node's representation by aggregating its representation and its ego network/connectivity representation using the relation: $e_h^{(1)} = f^{(1)}_{\mathcal{G}}(e_h, e_{\mathcal{N}_h})$, where aggregator $f^{(1)}_{\mathcal{G}}$ is defined as follows:
\begin{displaymath}
f^{(1)}_{\mathcal{G}} = \sigma\Big(W^{(1)}_1(e_h+e_{\mathcal{N}_h})\Big)+\sigma\Big(W^{(1)}_2(e_h\odot e_{\mathcal{N}_h})\Big).
\end{displaymath}
Here $\sigma$ is the Leaky ReLU activation function and $W^{(1)}_1, W^{(1)}_2 \in \mathbb{R}^{d^{(1)}} \times \mathbb{R}^d$ are trainable weight matrices. We follow a similar intuition for residual connections by aggregating information through the sum of two representations $e_h$ and $e_{\mathcal{N}_h}$ and retain a copy of input by using the identity transformation. Note that $\odot$ is the element-wise product that captures feature interaction between $e_h$ and $e_{\mathcal{N}_h}$, and ensures richer propagation of information from similar nodes.

For higher-order propagation, we stack more propagation layers to cascade information from higher-order neighbors. The $l$-th step node representation can be formulated as:
\begin{displaymath}
e_h^{(l)} = f^{(l)}_{\mathcal{G}}(e_h^{(l-1)}, e^{(l-1)}_{\mathcal{N}_h}),
\end{displaymath}
where the information cascaded from $l-1$-th ego network is defined as:
\begin{displaymath}
e_{\mathcal{N}_h}^{(l-1)} = \sum_{(h,r,t)\in \mathcal{N}_h} \pi'(h,r,t)e_t^{(l-1)}.
\end{displaymath}
Using this embedding propagation mechanism to stack $L$ layers, the higher-order connectivities can be captured in the node representation. Finally, we concatenate these representations into one vector to get the final representation of the node.
\begin{displaymath}e_h^* = [e_h^{(0)} \mathbin\Vert \cdots \mathbin\Vert \textrm{ } e_h^{(L)}]  \in \mathbb{R}^q,
\end{displaymath}
where $e_h^{(0)} := e_h$ and $q = d + d^{(1)} + \cdots + d^{(L)}$. The different layers of knowledge graph attention module described above are shown in Figure \ref{katrec} (left).

\subsection{Dynamic User Interaction Module}
To capture the sequential patterns among successive items that a user has interacted with, we use the Bidirectional Encoder Representations from Transformers (BERT) architecture \citep{vaswani2017attention, devlin2018bert}. In our context, BERT uses the historical item sequence $\mathcal{S}^u$ corresponding to user $u$ and aims to predict the item that the user is interested in the next time step $T+1$. BERT models $M$ bidirectional transformer layers and revises each item's representation at each layer by exchanging information across all positions at the previous layer. Our key contribution is that we embed the higher-order connectivity of the item relations discussed in subsection \ref{subsec:gat} into the BERT module to capture user preferences under a more informative context. Below, we briefly discuss the self-attention structure used in the BERT module of KATRec.

\subsubsection{Embedding Layer}
{Positional Embedding:}
Since the self-attention mechanism doesn't include any recurrent or convolutional blocks, it cannot be aware of items' position embeddings. So, we incorporate a pre-determined positional embedding $P \in \mathbb{R}^{T \times q}$ into the input embedding:
\begin{displaymath}
     v^{(0)}_i = e^*_i +  p_i,
\end{displaymath}
where $ v^{(0)}_i$ computes the input representation of items at each position index $i = 1,\cdots,T$. Concatenating embedding of $T$ items in a user sequence, $v^{(0)}_i \in \mathbb{R}^{q}$, results in $V^{(0)} \in \mathbb{R}^{T \times q}$. Next, this positional embedding matrix is provided as an input to the first transformer layer. The transformer layer contains two sublayers, a multi-head self-attention sublayer, and a position-wise feed-forward network. In the following, we describe each of these sublayers briefly.

\subsubsection{Transformer Layer}
Multi-Head Self-Attention: The attention mechanism helps the model capture dependencies between each pair of items at any distance in the input sequence, across multiple subspace representations simultaneously. To learn information in different representation subspaces, we use multi-head attention \citep{devlin2018bert}. First, we linearly project ${V}^{(0)}$ into $k$ subspaces using different learnable projections and then apply attention function on each in parallel to create $k$ heads $H_1, \cdots, H_k$ as follows:

\begin{displaymath}H_j = \text{Attention}(V^{(0)} W_j^\mathcal{Q}, V^{(0)} W_j^\mathcal{K}, V^{(0)} W_j^\mathcal{V}),
\end{displaymath}

where $W_j^\mathcal{Q}$, $W_j^\mathcal{K}$ and $W_j^\mathcal{V}$ are all $\mathbb{R}^{q\times q/k}$ learnable projection matrices corresponding to head index $j = 1, \cdots, k$. The attention function is a scaled dot-product defined as $\text{Attention}(\mathcal{Q},\mathcal{K},\mathcal{V}) = \text{softmax}\Big(\frac{\mathcal{Q}\mathcal{K}^\top}{\sqrt{q/k}}\Big)\mathcal{V}$. To compute the multi-head attention output, we concatenate these $k$ heads and then project it as follows:
 \begin{displaymath}\text{MH}({V^{(0)}}) = [{H}_1 \mathbin\Vert {H}_2\mathbin\Vert \cdots \mathbin\Vert {H}_k]W^O,
\end{displaymath}
where the learnable projection matrix $W^O \in \mathbb{R}^{q\times q}$. Next, we input $\text{MH}(V^{(0)})$ to a feed-forward sublayer described below.\\

\noindent Position-wise Feed-Forward Network: Since the self-attention sub-layer mainly leverages linear transformations, we need another sub-layer to implement non-linearity and also capture interactions between different dimensions. We use a position-wise feed-forward network on the outputs of the multi-head self-attention sublayer at each position $i = 1, \cdots, T$.
\begin{displaymath}
\text{PFF}(V^{(0)}) = [FF(v_1^{(0)})^\top \mathbin\Vert \cdots \mathbin\Vert FF(v_{T}^{(0)})^\top]
\end{displaymath}
where $FF(x) = \text{GELU}(xW^1+b^1)W^2+b^2$ uses the GELU non-linearity, and $W^1, {W^2}^\top \in \mathbb{R}^{q\times 4q}$, $b^1 \in \mathbb{R}^{4q}$ and $b^2 \in \mathbb{R}^{q}$ are learnable and shared at different positions.

Furthermore, we apply dropout to the output of both multi-head self-attention and position-wise feed-forward sub-layers. Due to the architecture depth, we also add residual connections to capture item-item interactions better. Finally, we apply layer normalization (LN) to  stabilize and accelerate training. These operations can be summarized as $\textrm{LN}(x + \textrm{Dropout}(\textrm{sublayer}(x)))$. In order to learn a more complex representation of items in the sequence, we stack $M$ Transformer layers as illustrated in Figure \ref{katrec}. Again, we note that the above choices are informed by the original BERT architecture for language modeling.

\subsubsection{Output Layer}
Item Co-occurrence Modeling: Capturing pairwise item relations plays an important role in the effectiveness of the recommendation systems and also allows some degree of interpretability. To include item-item relations in item embeddings, we concatenate item embeddings learned by the sequential module and the knowledge graph. 
\begin{displaymath}
    \hat{E} = \sigma((V^*\mathbin\Vert E^*)\hat{W} + {\hat{b}}),
\end{displaymath}
where $\hat{E}\in \mathbb{R}^{|\mathcal{I}|\times q}$ is the set of final KATRec embeddings for item set $\mathcal{I}$ and $\hat{W} \in \mathbb{R}^{2q\times q}$ and $\hat{b} \in \mathbb{R}^{q}$ are learnable parameters. $V^*$ is the embedding matrix of items learned by sequential module, and $E^*$ is the items embedding table learned by the knowledge graph. Finally, the next item at time $T+1$ is predicted by:
\begin{displaymath}
P(\mathcal{I}) = \mbox{softmax}(\mbox{GELU}({v}^{(M)}_{T+1} {W}^P + {b}^P) {\hat{E}^\top} + {b}^O),
\end{displaymath}
where ${W}^P, {b}^P$, and ${b}^O$ are learnable parameters, ${\hat{E}} \in \mathbb{R}^{|\mathcal{I}| \times q}$ is the embedding matrix for item set $\mathcal{I}$, and $P(\cdot)$ is the KATRec model's predicted distribution over the target items. $v^{(M)}_{T+1}$ is the hidden state of position $T+1$ after $M$ transformer layers, denoted by $v^*_{T+1}$ in figure \ref{katrec}. \\

\noindent Explicit User Modeling: Existing approaches provide personalized recommendation by modeling the user embedding either explicitly \citep{koren2008factorization, rendle2010factorizing, tang2018personalized} based on users' previous actions, or implicitly \citep{Hidasi2018, kang2018self, sun2019bert4rec} based on embeddings of the sequence of visited items by a user. KATRec belongs to the latter category as we predict the next item at time step $T+1$ by considering the hidden state embedding $v^{(M)}_{T+1}$. 

As an aside, we also considered explicit user behavior by incorporating user embeddings learned from the knowledge graph into the user's hidden state via concatenation: $[e^*_u \mathbin\Vert v^{(M)}_{T+1}]$, where $e^*_u$ is the embedding of user $u$ in the knowledge graph. Although this concatenation seems promising, we empirically did not observe improvement in the model's performance. This could be potentially because the model learned users' embedding very well by considering the sequence of interacted items. 

\subsection{Optimization}

The loss function of KATRec contains the TransR objective along with a regularizer. In particular, we use the TransR to train the entity embeddings. The objective function can be minimized by discriminating between valid and invalid triplets in the collaborative knowledge graph:
\begin{equation}\label{loss_CKG}
\mathcal{L}_{\mathcal{G}} = \sum_{(h, r, t, t')} -\text{ln} \ \sigma \Big(s(h, r, t')-s(h, r, t)\Big) + \lambda \lVert \Theta\rVert_2^2,\end{equation}
where the sum is over all valid $(h, r, t) \in \mathcal{G}$ and invalid $(h, r, t') \notin \mathcal{G}$ triplets in the knowledge graph $\mathcal{G}$, $\sigma(\cdot)$ is the sigmoid function, and $\lambda \lVert \Theta\rVert_2^2$ represents $\ell_2$ regularization. Similar to \cite{devlin2018bert}, we  implement the \textit{Cloze task} training approach in addition to pairwise ranking loss in equation (\ref{loss_CKG}). This allows us to learn the parameters of the encoders in the transformer layers. In this approach, we randomly mask a portion of items in the input sequence $\mathcal{S}_u$ and try to predict them. The loss for each masked input $S'_u$ is given by:
\begin{displaymath}
    \mathcal{L}_{\mathcal{S}} = \frac{1}{|\mathcal{S}_u^m|} \sum_{i_m \in \mathcal{S}_u^m} - \log P(i_m = i^\star_m| \mathcal{S}'_u ),
\end{displaymath}
where $\mathcal{S}_u^m$ is the set of randomly masked items, $i^\star_m$ is the true item corresponding to the masked item $i_m$, and $P(.)$ is the predicted probability mass function over the target item. These two losses are jointly minimized over their respective parameters using standard first-order approaches (see the next section for details).

\section{Experiments}\label{experiment}

We evaluate our model on three real-world datasets, which are different in domains and have varying levels of sparsity\footnote{Code is available at \url{https://github.com/DanialTaheri/KATRec}}. We aim to answer the following questions in this section:\\
{Q1:} How does KATRec perform compared to the current state-of-the-art sequential recommendation methods?\\
{Q2:} How do different components of the model (viz., knowledge graph based attention mechanism, information aggregation, and pre-training) affect the performance of KATRec?\\
{Q3:} How KATRec's performance change in different settings and datasets?

Following prior work \cite{sun2019bert4rec, kang2018self}, we convert all numeric ratings to positive interaction with a value of $1$, which indicates that the user has interacted with the item. Then, we sort each user's interactions by timestamp to build her interaction sequence.  Similar to prior works, we split the user sequences into three parts. The test dataset includes the most recent item each user has interacted with ($S^u_{T+1}$), the validation dataset consists of the second most recent item interacted by each user ($S^u_{T}$), and the remaining items in the sequence belong to the training data. To construct knowledge graph aware attention, we first build the item knowledge graph for each dataset. We follow \cite{zhao2019kb4rec, KGAT19} to capture knowledge graph triplets by mapping items into \emph{freebase} entities. We include triplets with one-hop and two-hop neighbor entities and filter out entities with less than ten occurrences, and relations less than fifty occurrences. For $\mathcal{L}_{\mathcal{G}}$ in equation (\ref{loss_CKG}), we pair each observed triplet with a broken (unobserved) triplet.
\subsection{Datasets Description}
We evaluate our model on three real-world datasets, which are different in domain and sparsity. For most of the experiments, we only include users and items with at least ten interactions to ensure data quality \cite{KGAT19}. The statistics of the datasets described below are presented in Table~\ref{tab:data} for ease of reference:\\

\begin{table}[ht]
	\centering
	\caption{Statistics of datasets}\label{tab:data}
	\resizebox{\textwidth}{!}{%
	\begin{tabular}{lccccccc}
	\hlineB{2}
    \textbf{Datsets} & Users  & Items & Interactions & Entities & Relations & Triplets & Density \\
    \hline
Amazon-book & 70679 & 24915 & 846434 & 88572 & 39 & 2557746 &0.048\%\\
LastFM & 23566 & 48123 & 8057269 & 58266 & 9 & 464567 &0.7105\%\\
Yelp2018 & 45919 & 45538 & 1185068 & 90961 &42 & 1853704 &0.057\%\\
	\hlineB{2}
    \end{tabular}%
    }%
\end{table}

{Amazon-book}: Amazon review data is one of the popular datasets in the recommendation systems literature \cite{he2016ups}. The data has been categorized based on different product categories, and in this paper, we focus on the book category.\\
{LastFM}: This is a dataset about music listening patterns collected from the Last.fm online music platform \cite{schedl2016lfm}. In this dataset, tracks are viewed as items, and we consider a subset of the dataset from January 2014 to August 2014.\\
{Yelp2018}: This dataset is adopted from the Yelp 2018 recommendation system challenge. In this dataset, local restaurants and bars are represented as items. 

\noindent We calculate each dataset's density based on the number of interactions, users, and items as $\frac{|Interactions|}{|Users| \cdot |Items|}$. Therefore, larger values in the density column represent datasets with more interactions (per user and item). Table \ref{tab:data} lists specific properties and the density of each dataset. It also highlights that LastFM has a substantially lesser number of relations and is significantly denser than others.

\subsection{Experimental Settings}
\subsubsection{Evaluation Metrics}
We use two common top-K metrics to evaluate the performance of our model. Hit@K and NDCG@K count the fraction of the time the ground truth item is among the top $K$ recommendations, without and with defining a position-aware weight respectively. Mean Average Precision (MAP) is also a ranked precision metric over all users that emphasizes correct predictions at the top of the list with a position-aware weight. Given a list of top $K$ predicted items $\hat S^u_{1:K}$ for a representative user $u$, and the ground truth last item $S^u_{T+1}$, Hit@K, NDCG@K, and MAP  are computed as:
\begin{align*}
    Hit@K &= \frac{1}{|\mathcal{U}|}\sum_{u \in \mathcal{U}}|\hat S^u_{1:K} \bigcap S^u_{T+1}|,\\
    NDCG@K &= \frac{1}{|\mathcal{U}|}\sum_{u \in \mathcal{U}}\sum_{i=1}^K\frac{rel_i}{\log_2(i+1)},\\
    MAP &= \frac{1}{|\mathcal{U}|}\sum_{u \in \mathcal{U}}\sum_{i=1}^{|\hat S^u|} Prec@i \times rel_i,
\end{align*}
where $rel_i = 1$ if the $i^{th}$ item in $\hat S^u_{1:K}$ equals $S^u_{T+1}$, and $Prec@i$ is the precision (fraction of  recommendations that end up being correct) up till position $i$. Following the works of \cite{kang2018self, sun2019bert4rec}, we randomly sample $100$ negative items for each user besides the ground-truth item. We report the average of the metrics Hit@K, NDCG@K over all users.

\subsubsection{Baselines}
To compare the performance of our model with others, we consider the following competitive baselines:
\begin{itemize}
    \item GRU4Rec \citep{hidasi2015session}: It implements a session-based recommendation model based on RNNs. We consider each user's sequence as a session.
    \item GRU4Re$c^{++}$ \citep{Hidasi2018}: It modifies the way GRU4Rec is optimized by implementing a new loss function and a new sampling approach.
    \item SASRec \citep{kang2018self}: It uses a self-attention mechanism with a left-to-right Transformer to improve the capturing of useful patterns in user sequences.
    \item BERT4Rec \citep{sun2019bert4rec}: This is a recent state-of-the-art sequential recommendation model that adapts the bidirectional Transformers language model architecture to learn the temporal behavior of users.
\end{itemize}

\subsubsection{Parameters for Models}

We implement KATRec with Tensorflow (version 2.2.0). All parameters are initialized using a truncated normal distribution in the interval $[-0.02, 0.02]$. We use the Adam optimiser \citep{kingma2014adam} with learning rate $10^{-4}$ that decays linearly, $\beta_1 = 0.9$, $\beta_2 = 0.999$, and a weight decay of $0.01$. We fix the maximum sequence length proportional to the average sequence length in the dataset, i.e., $50, 50$, and $200$ for Amazon-book, Yelp, and LastFM respectively. We set the dimension of the hidden fully connected layers of KATRec to be $128$. We propagate neighbors' information up to three levels into each entity's embedding with hidden dimensions $32$, $16$, and $16$. Finally, we set the embedding of entities in the knowledge graph to be $64$. 

We search for hyperparameters to select the best parameters for different baselines. These include changing embedding size from $\{8, 16, 32, 64, 128\}$ and the regularization hyperparameter $\lambda$ across $\{0.1, 0.05, 0.01, 0.005, 0.001, 0.0001\}$. We use the optimization schemes and parameters suggested by the authors whenever possible. The models are trained on a single GeForce GTX 1080Ti GPU.

\subsubsection{Performance Comparison}

Table \ref{tab:PerformanceComparison} shows the recommendation performance of KATRec and baselines. We do not include Hit@1 since Hit@1 and NDCG@1 are equivalent. Since we have a single ground-truth, Hit@K is equivalent to Recall@K, and it is proportional to Precision@K. \emph{We observe that KATRec provides improved relative recommendation performance over all alternatives by 6.29\% and 3.82\% in Hit and 7.15\% and 4.64\% in NDCG on average, respectively on the Amazon-book and Yelp2018 datasets.}

\begin{table}[ht]
  \caption{KATRec versus baselines over three datasets. The improvement percentage compares KATRec versus the next-best alternative.}
  \label{tab:PerformanceComparison}
  \begin{tabular}{clcccccc}
    \hlineB{2}
    Datasets  & Metrics &GRU &$GRU^{++}$ &SASRec &BERT &KATRec &Improv. \\
    \hline
    &NDCG@1 &0.3485 &0.3464 &0.3749 &\underline{0.4344} &\textbf{0.4706} &8.33\%\\
    &NDCG@5 &0.4404 &0.4358 &0.5267 &\underline{0.5715} &\textbf{0.6110} &6.91\%\\
    {Amazon}&NDCG@10 &0.4598 &0.4574 &0.5600 &\underline{0.6022} &\textbf{0.6401} &6.2\%\\
    &Hit@5 &0.5202 &0.5148 &0.6594 &\underline{0.6910} &\textbf{0.7321} &5.94\%\\
    &Hit@10 &0.58 &0.5814 &0.7621  &\underline{0.7856} &\textbf{0.8217} &4.6\%\\
    &MAP &0.42 &0.4259 &0.5065 &\underline{0.5539} &\textbf{0.5907} &6.64\%\\
    \hline
    & NDCG@1 &0.3646 &0.3523 &\underline{0.6771} &0.6339 &\textbf{0.6931} &2.36\%\\
    &NDCG@5 &0.4648& 0.4448 &\textbf{0.7765} &0.7606 &\underline{0.7725}  &-0.51\%\\
    {LastFM}&NDCG@10 &0.4881 &0.4674 &\textbf{0.7930} &0.7786 &\underline{0.7911} &-0.24\%\\
    &Hit@5 &0.5531 &0.5263 &\textbf{0.8600} &0.8281 &\underline{0.8426}  &-2.06\%\\
    &Hit@10 &0.6249 &0.5958 &\textbf{0.9105}  &0.8836 &\underline{0.9001} &-1.15\%\\
    &MAP &0.4577&0.4357&\underline{0.7598} & 0.7509  &\textbf{0.7618} &0.26\%\\
    \hline
    &NDCG@1 &0.3946&0.4148&0.3723 &\underline{0.4149} &\textbf{0.4405} &6.17\%\\
    &NDCG@5 &0.5041 &0.5143 &0.5703 &\underline{0.6039} &\textbf{0.629} &4.15\%\\
    {Yelp2018}&NDCG@10 &0.5278 &0.5395 &0.6068 &\underline{0.6400} &\textbf{0.663} &3.6\%\\
    &Hit@5 &0.5991 &0.6021 &0.7434 &\underline{0.7690} &\textbf{0.7927} &3.08\%\\
    &Hit@10 &0.6721 &0.68 &0.8551 &\underline{0.8796} &\textbf{0.899} &2.2\%\\\
    &MAP &0.49&0.515&0.5351 &\underline{0.5706} &\textbf{0.5946} &4.21\%\\
    \hlineB{2}
\end{tabular}
\end{table}

In particular, KATRec consistently outperforms BERT4Rec on Amazon-book and Yelp2018 datasets, which shows the importance of modeling item-item and user-item relations. KATRec achieves a considerable performance improvement in Amazon-book, while the improvement in Yelp is relatively small. This observation can be attributed to the difference in the sparsity of these two datasets. The importance of the impact of data density and number of relations on KATRec's performance is highlighted explicitly with the LastFM dataset, which we discuss in further detail in the ablation study that follows.

\subsubsection{Ablation Study}
In this section, we analyze variants of KATRec to understand the impact of different components on model performance. The variations are as follows: 
(1) \textit{No Attention}: we remove the attention mechanism in the knowledge graph and allocate equal weights to each entity's neighbors. 
(2) \textit{Level-1}: we decrease the level of information that can propagate from the neighbors to a node, and study the impact of only using immediate neighbors to improve node embeddings.
(3) \textit{Connection}: While there exists a connection between two encoder modules in our model, we consider the setting where both modules train independently using learnable embeddings.
(4) \textit{No Pretraining}: We forgo the pre-trained embeddings of entities in the knowledge graph.
(5) \textit{Concat}:  We remove a part of our model that deals with item co-occurrence, and only consider the item embedding vector that results from the sequential module.

Results are shown in Table \ref{tab:Ablation}. We observe that incorporating the side information in the bidirectional encoder module during the training results in better parameter learning as shown in column \textit{Connection}. Furthermore, the results of making the \textit{NoPretrain} choice in the knowledge graph show that incorporating pre-trained embeddings increases the performance of the sequential recommendation model. The attention mechanism between neighbors in the knowledge graph increases the recommendation's performance. However, this increase is not substantial. We also observe that multiple layer information propagation in an item's embedding plays an important role in making next item recommendations. Also, as expected, propagating the first layer's information in the knowledge graph has a higher impact, and this impact decreases when we incorporate higher level of connections. Finally, the results of the \textit{Concat} choice shows that incorporating non-linearity in the final layer is beneficial for learning item embeddings, especially by combining features learned through the knowledge graph and the bidirectional encoder.
\begin{table}[ht]
    \centering
    \caption{Ablation study of design choices in KATRec using three datasets.}
    \label{tab:Ablation}
    \begin{tabular}{clcccccc}
    \hlineB{2}
     Datasets &Metrics &KATRec &NoAtten. &Level-1 &Connect. &NoPretrain &Concat.\\
     \hline
     {Amazon-}&NDCG@10 &0.6401 &0.6371 &0.6386&0.621 &0.6306 &0.6318 \\
     {book}&Hit@10 &0.8217 &0.8178 &0.8195 &0.801  &0.8092 &0.8142 \\
     \hline
     {LastFM}&NDCG@10 &0.7911 &0.7836 & 0.7853 &0.763 &0.7587 &0.7855 \\
            &HIT@10 &0.9001 &0.8967 &0.8957 &0.8796 &0.8752 &0.8908 \\
    \hline
     {Yelp2018}&NDCG@10 &0.663 &0.6567 &0.6546 &0.6458 &0.6359 &0.6515 \\
        &HIT@10 &0.899 &0.8954 &0.8929 &0.885 &0.8696 &0.8881\\
    \hlineB{2}
    \end{tabular}
\end{table}

\begin{figure}[h]
\begin{subfigure}{0.325\textwidth}
\centering
\includegraphics[width=\textwidth]{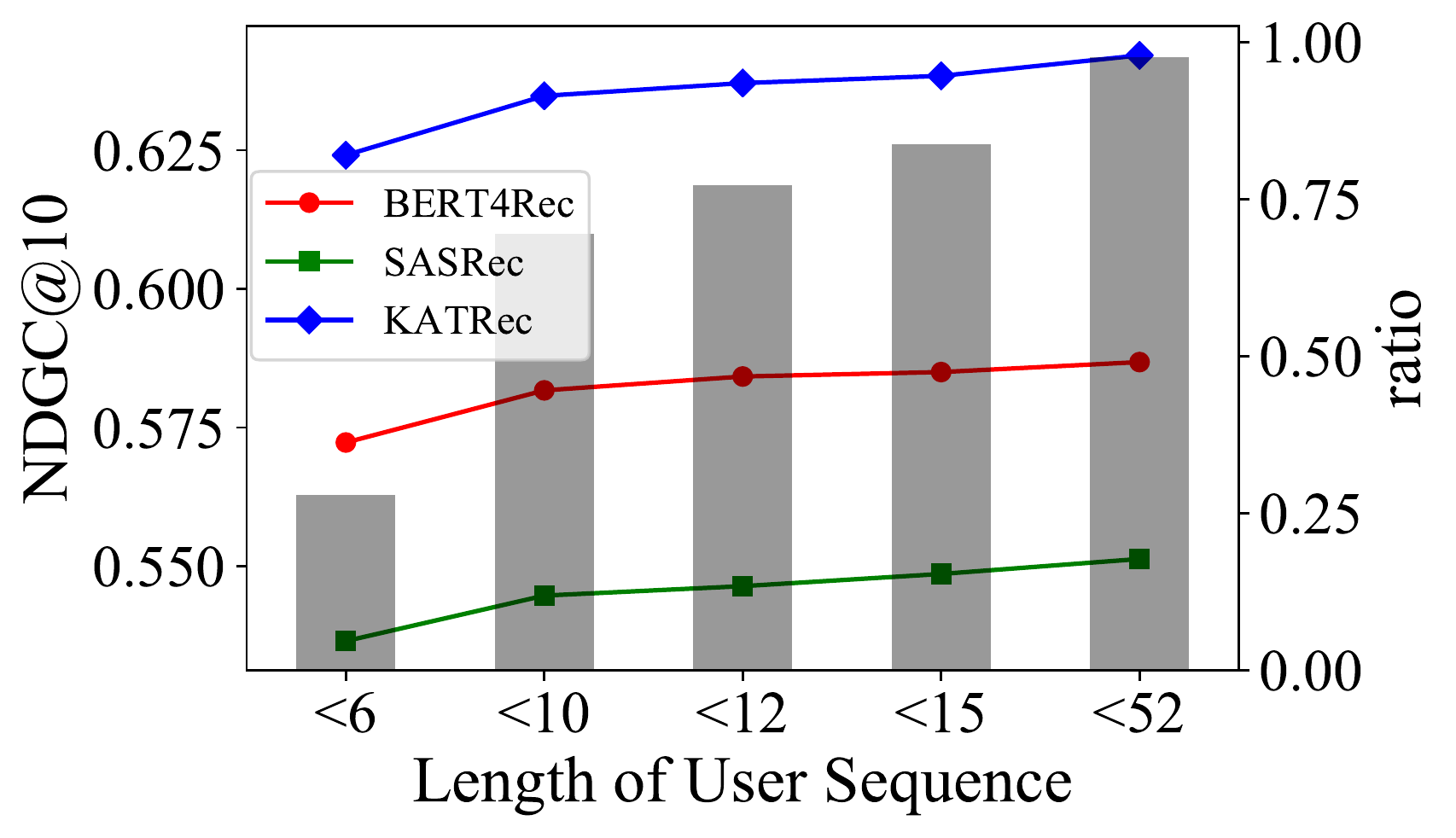}
\subcaption{Amazon-book}
\end{subfigure}
\begin{subfigure}{0.33\textwidth}
\centering
\includegraphics[width=\textwidth]{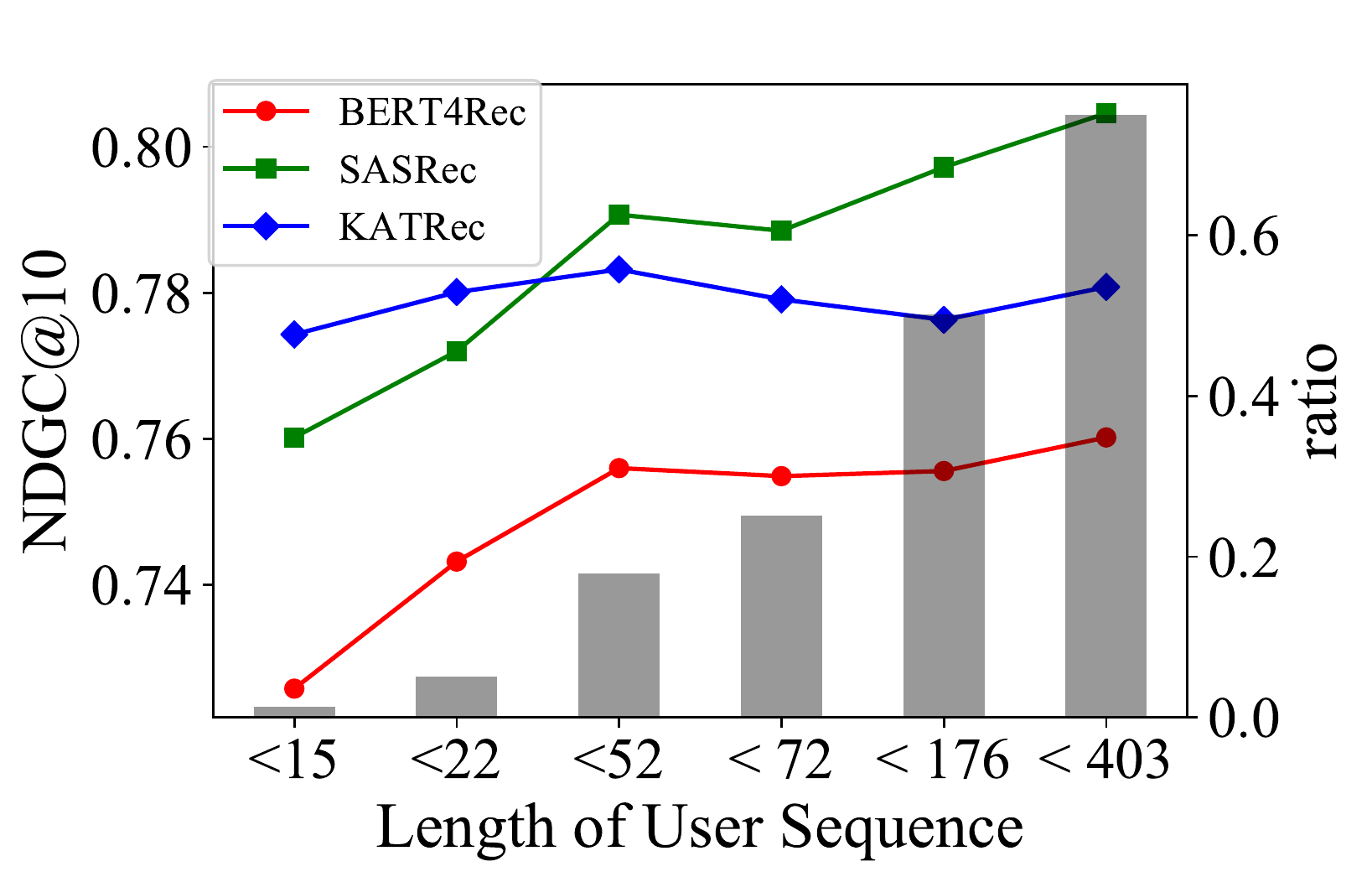}
\subcaption{LastFm}
\end{subfigure}
\begin{subfigure}{0.325\textwidth}
\centering
\includegraphics[width=\textwidth]{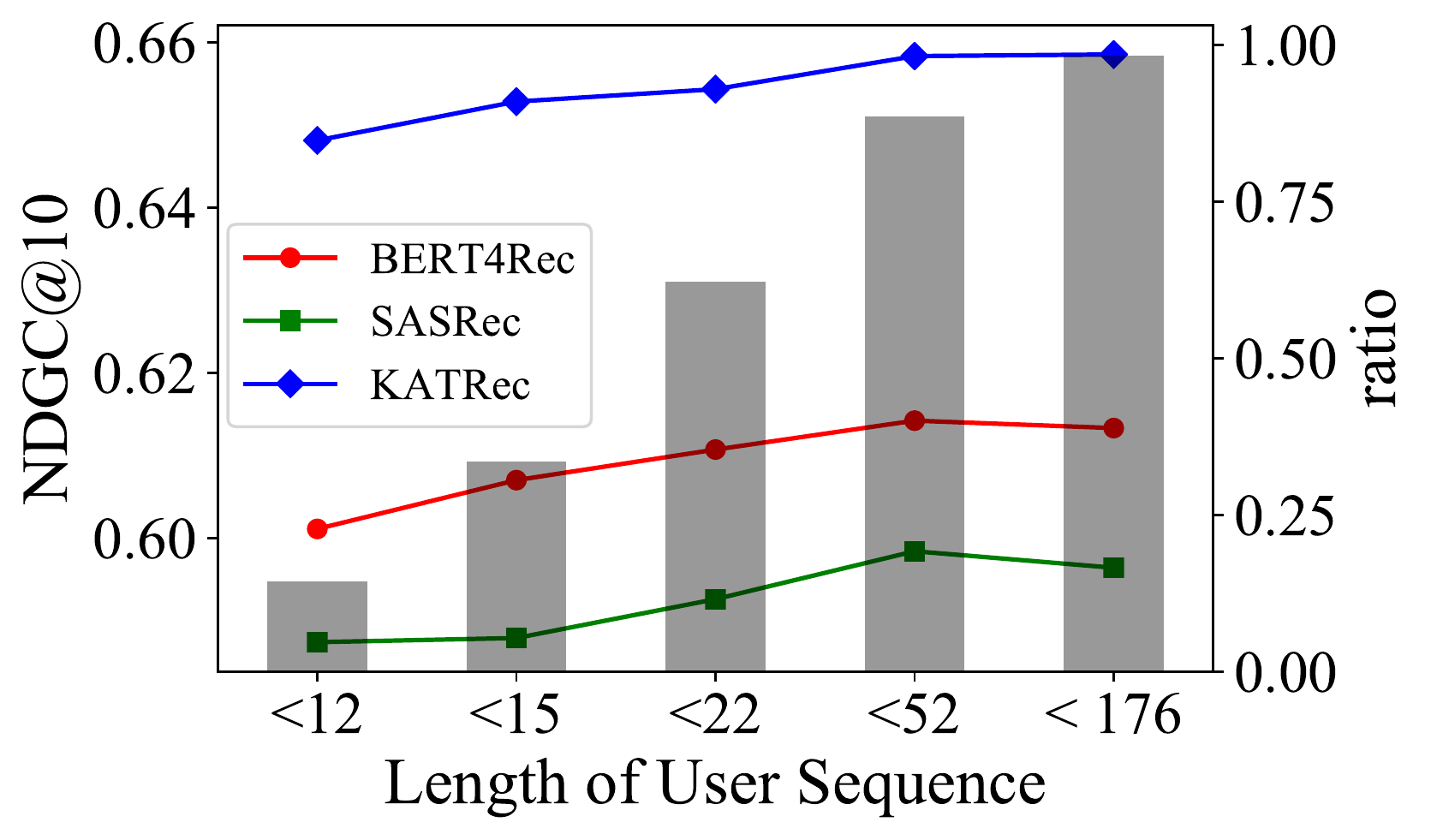}
\subcaption{Yelp}
\end{subfigure}
\caption{Performance comparison of models across users with different sequence lengths on the Amazon-book, LastFM and Yelp2018 datasets.}\label{seq-analysis}
\end{figure}

We also study the recommendation performance of KATRec for users with different sequence lengths and compare it with competitive baselines. The intuition here is that the use of side information can compensate for potentially sparse user item interactions. In Figure \ref{seq-analysis}, we illustrate the percentage of users with varying sizes of item sequences associated with them in each dataset, and report each method's performance for each of the resulting user groups. KATRec outperforms two other competitive baselines for all user groups for Yelp2018 and Amazon-book datasets. Results for LastFM show that KATRec provides a reasonably robust recommendation performance across user groups, while the performance of the baselines is more sensitive to the user sequence length. This robustness of KATRec can be attributed to the item-item and user-item relationships that are explicitly learned. Our model performs relatively better for users with small sequence lengths. However, SASRec marginally outperforms KATRec for users with long sequence lengths. This can be attributed to the high number of interactions and low number of relations in the LastFM dataset, which indicates that the impact of side information on dense datasets may not be significant enough.

\subsubsection{Visualizing attention weights} 
\begin{figure}[h]
\centering
\captionsetup[subfigure]{justification=centering}
\begin{subfigure}{0.24\textwidth}
\centering
\includegraphics[width=\textwidth]{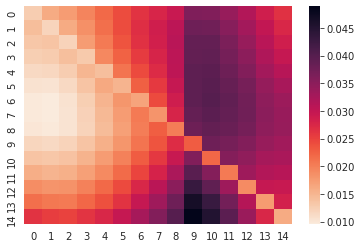}
\subcaption{KATRec,\\ Layer 1, head 1}
\end{subfigure}
\begin{subfigure}{0.24\textwidth}
\centering
\includegraphics[width=\textwidth]{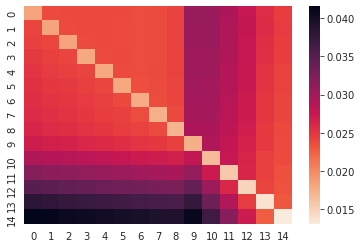}
\subcaption{KATRec, Layer 1,\\ head 1, w/o PE}
\end{subfigure}
\begin{subfigure}{0.24\textwidth}
\centering
\includegraphics[width=\textwidth]{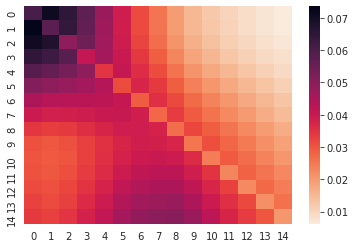}
\subcaption{KATRec,\\ Layer 2, head 2}
\end{subfigure}
\begin{subfigure}{0.24\textwidth}
\centering
\includegraphics[width=\textwidth]{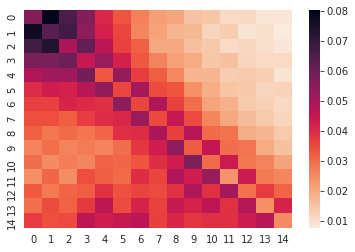}
\subcaption{BERT4Rec,\\ Layer 2, head 2}
\end{subfigure}
\caption{Average attention weights on positions (x-axis) at different time (y-axis) on Yelp}
\label{Attention_weights}
\end{figure}

This section visualizes the attention weights related to items and positions to find a meaningful pattern and discuss their differences with the BERT model's attention weights. Figure \ref{Attention_weights} shows the heatmaps of the average attention weights on the last 15 items of the sequences in the test dataset of Yelp2018. In order to calculate the accurate average weights, we do not incorporate weights of padded items in sequences shorter than 15 items. Comparing heatmaps $(a)$ and $(b)$ shows the impact of positional embeddings (PE). In particular, heatmap $(a)$ illustrates how positional embeddings results in items attending more on recent items. Heatmaps $(a)$ and $(c)$ points out how items in various heads and layers focus on different parts of the sequence at both the right and left sides. To compare attentions in BERT4Rec and KATRec, we analyze weights in the final layer as it is directly connected to the output layer and plays an important role in the prediction (heatmaps $(d)$ and $(c)$). The comparison indicates that while BERT4Rec inclines to focus more on the recent items due to the sparsity of the dataset, KATRec tends to attend on less recent items due to incorporating side information through the knowledge graph. This behavior is similar to the attention weights of self-attention blocks in dense datasets in \cite{kang2018self}. 

\subsubsection{Attention Weight Case Study}
\begin{figure}[h]
\centering
\captionsetup[subfigure]{justification=centering}
\begin{subfigure}{0.4\textwidth}
\centering
\includegraphics[width=\textwidth]{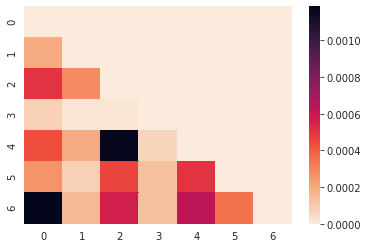}
\subcaption{Co-occurence information}\label{cooccurence_Parta}
\end{subfigure}
\begin{subfigure}{0.4\textwidth}
\centering
\includegraphics[width=\textwidth]{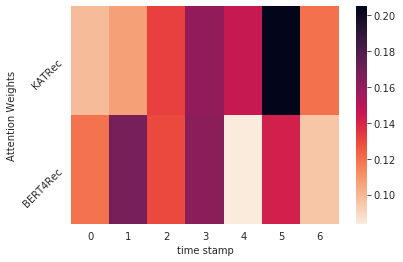}
\subcaption{Attention weights}\label{att_Partb}
\end{subfigure}
\caption{Attention heatmap comparison of a random user in Yelp2018 dataset. \ref{cooccurence_Parta} shows the co-occurrence ratio, which is the co-occurrence frequency of each pair of six items among all users' sequences. \ref{att_Partb} compares these items' attention weight at the last position in KATRec and BERT4Rec.}
\end{figure}

Figure \ref{cooccurence_Parta} illustrates the co-occurrence ratio between six items in user sequences, computed by the average number of times that a pair of items appeared simultaneously in users' sequence. Figure \ref{att_Partb} compares weights of the final attention layer in KATRec and Bert4Rec between the last item (item $6$) and the rest of the items in a user sequence. Figure \ref{cooccurence_Parta} shows that item $6$ has a high co-occurrence value with items $0, 2, 4$, and $5$. Figure \ref{att_Partb} confirms that KATRec places a higher attention values for items $4$ and $5$ and lower values for items with low co-occurrence value, i.e., time $1$ and $3$. However, we observe that BERT4Rec considers a higher weight between item $6$ and $0$ which is more aligned to their co-occurrence value.

\section{Conclusion and Future Work}\label{conclusion}
Designing robust deep neural network architectures that produce quality recommendations is challenging for several reasons. A couple of these challenges were addressed in this work, namely leveraging of side information and getting around data sparsity. In particular, we proposed incorporating item side information to alleviate both these shortcomings while making recommendations. This information is readily available in many real-world applications. 

Our work introduces a novel neural network structure that leverages collaborative knowledge graphs to improve the representations of items in a sequential recommendation system setup. Empirical results are provided to illustrate the benefit via multiple evaluation metrics: the proposed solution is compared against multiple state-of-the-art sequential recommendation systems on three different datasets. Similar to the way we included item metadata in building a more performant recommendation system, further research in incorporating user metadata can be undertaken.

\vskip 0.2in
\bibliography{KATRecbib}

\end{document}